\documentclass[a4paper,twocolumn,11pt,unpublished]{quantumarticle}
\pdfoutput=1
\usepackage[utf8]{inputenc}
\usepackage[english]{babel}
\usepackage[T1]{fontenc}
\usepackage{amsmath,amssymb}
\usepackage{hyperref}

\usepackage{tikz}
\usepackage{lipsum}

\newcommand{\Mod}[1]{\mathrm{mod}\, #1}
\DeclareMathOperator{\Si}{Si}\DeclareMathOperator{\sinc}{sinc}
\DeclareMathOperator{\lcm}{lcm}

\begin{document}

\title{Benchmarks for quantum computers from Shor's algorithm}

\author{E. D. Davis}
\email{david.davis@spu.ac.za}
\orcid{0000-0001-8957-6687}
\affiliation{Department of Physical and Earth Sciences, 
 Sol Plaatje University, Private Bag X5008, Kimberley 8300, South Africa}
\maketitle

\begin{abstract}  
Properties of Shor's algorithm and the related 
period-finding algorithm could serve as benchmarks for the operation of a quantum computer. 
Distinctive universal behaviour is expected for the probability for success of the period-finding 
algorithm as the input quantum register is increased through its critical size of  
$\mathfrak{m}_0=\lceil 2\log_2 r\rceil$ qubits (where $r$ is the period sought). Use of quadratic 
non-residues permits unequivocal predictions to be made about the outcome of the factoring algorithm. 
\end{abstract}

\section{Introduction}

The rapid development of quantum technologies over the last decade has prompted much research
on techniques for checking that the carefully engineered quantum devices are, in fact, functioning
properly \cite{GKK19,EHW20,KR21}. 
In the survey of the field presented in \cite{EHW20}, a distinction is drawn between certification (or 
verification) and benchmarking. Certification is taken to be confined to the assessment the accuracy 
of the output, whereas benchmarking protocols are any more general measures of performance.
Clearly, in this parlance, Shor's factoring algorithm and the associated period-finding algorithm can 
be used for verification as the validity of their output is easily checked, but the ancillary properties 
mentioned in the abstract and discussed below could serve as the basis for benchmarks that    
test circuits as a whole.

In Shor's version of his factoring algorithm \cite{Sh97}, a square-free semiprime $N$ is factored by 
first using a quantum computer to determine the order $r$ of an arbitrary element  
$b\in (\mathbb{Z}/N\mathbb{Z})^\times$, or, more colloquially, the period of $b^x \Mod{N}$, 
and then using a classical computer to calculate the candidate $f_b =\gcd(b^{r/2}-1,N)$
for a non-trivial factor of $N$. (Unless specified otherwise, the notation of \cite{CvD10} is adopted.)

\begin{figure}[t]
  \centering
  \includegraphics[width=0.45\textwidth]{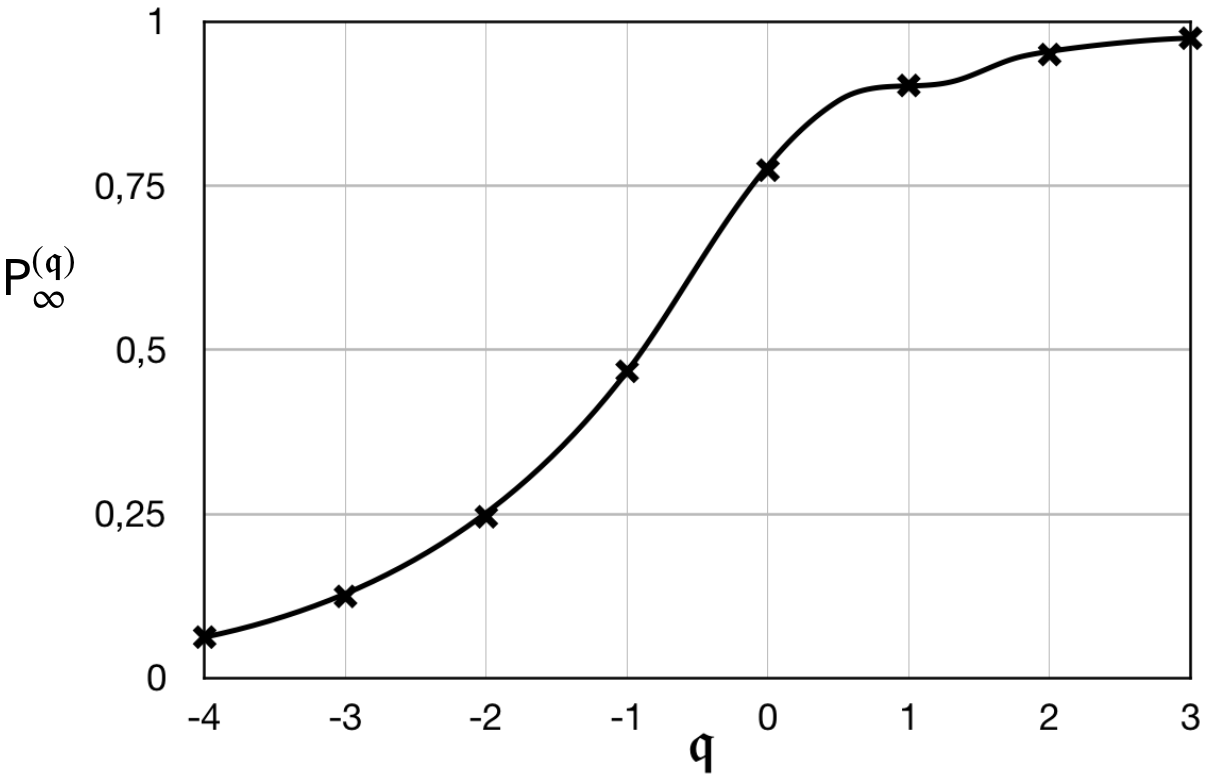}
  \caption{A plot of the approximation $\mathsf{P}_\infty^{(\mathfrak{q})}$ in \eqref{eq:Pq}  to the probability of success of the period-finding
  algorithm versus the increment $\mathfrak{q}$ of the input register size from its critical size of $\mathfrak{m}_0=\lceil 2\log_2r\rceil$ qubits. 
}
  \label{fig:figure1}
\end{figure}

It is the dependence on the period $r$ of the probability for success of the period-finding 
algorithm which is relevant to benchmarking.
The nature of this dependence can be established by considering the sequence of input register sizes
$\{ \mathfrak{m}_{\mathfrak{q} } \}$, where $\mathfrak{m}_{\mathfrak{q} }$ is the \emph{smallest\/} 
positive integer such that
\begin{equation}\label{eq:mqdef}
 2^{ \mathfrak{m}_\mathfrak{q} } > 2^\mathfrak{q} r^2 ,
\end{equation}
and $\mathfrak{q}$ is \emph{any\/} integer consistent with the requirement that the input register is large enough
for the quantum Fourier transform of its state to yield information on $r$, i.e. $2^{\mathfrak{m}_\mathfrak{q}}\ge 2r$
from the $k=r-1$ version of the first inequality in \eqref{eq:constraint}, or
$\mathfrak{q}\ge\mathfrak{q}_{\min} = \lceil\log_2 r\rceil- \lfloor 2\log_2 r \rfloor$.

A systematic asymptotic analysis implies that, when $r\gg 1$,  a useful 
approximation to the probability of finding a divisor of $r$ (or $r$ itself) with an 
input register of $\mathfrak{m}_{\mathfrak{q} }$ qubits  is
\begin{equation}\label{eq:Pq}
\mathsf{P}^{(\mathfrak{q})}_\infty =
  \frac{2}{\pi}\Si(2^\mathfrak{q}\pi) -  \Bigl( \frac{\sin 2^{\mathfrak{q}-1}\pi }{2^{\mathfrak{q}/2-1}\pi}\Bigr)^2 
\end{equation}
($\Si(x)$ is the sine integral of Eq.~6.2.9 in \cite{DLMF}).
The monotonically increasing sequence 
$\{ \mathsf{P}^{(\mathfrak{q})}_\infty \}$ is depicted in Fig.~\ref{fig:figure1}. The interpolating curve is obtained with
the same function used to evaluate the $\mathsf{P}^{(\mathfrak{q})}_\infty$'s. As the derivation of section~\ref{sc:BW}
and \eqref{eq:ro1} make clear, the result in \eqref{eq:Pq} pertains to the determination of the period of 
\emph{any\/} function of the integers $\mathbb{Z}$ provided it is one-to-one within a period, and
the period $r$ is not a divisor of $2^{\mathfrak{m}_\mathfrak{q}}$. When $r$ is a power of two (and, hence, a
divisor of $2^{\mathfrak{m}_\mathfrak{q}}$), the probability of success is independent of $\mathfrak{q}$ [see \eqref{eq:ro1}].

The register of $\mathfrak{m}_0=\lceil 2\log_2r\rceil$ qubits
is distinguished by the fact that it is, generically, the smallest for which $\mathsf{P}^{(\mathfrak{q})}_\infty$ exceeds 50\%.
Among the inferences that one can draw from Fig.~\ref{fig:figure1} is that \emph{the magnitude of the 
probability of success for the period-finding algorithm amounts to a fingerprint of the critical input register}. 
Statistics on the success or failure of runs for  $\mathfrak{m}_0$ qubits would permit empirical determination of 
the corresponding probability 
of success of the period-finding algorithm; unambiguous comparison of the probability found with 
$\mathsf{P}^{(\mathfrak{q=0})}_\infty$ should be possible in view of its clear separation in value from the other 
$\mathsf{P}^{(\mathfrak{q})}_\infty$'s: this comparison constitutes the first benchmark,
applicable whenever
$r$ is not a divisor of $2^{\mathfrak{m}_0}$ (as is typically the case).

Corrections to $\mathsf{P}_\infty^{(\mathfrak{q})}$, considered in section \ref{sc:asymp}, do not invalidate any 
of the observations above. In fact, it is found that, for $\mathfrak{q}\ge0$, 
these corrections are negligible (see \eqref{eq:asympq}, \eqref{eq:asymp0} and Table \ref{tb:qc2}). 
The clear difference between the success probability for the critical input register and input registers of 
other sizes also survives (see Fig.~\ref{fig:figure3}).

Another opportunity for benchmarking is provided by
the choice of the element $b\in(\mathbb{Z}/N\mathbb{Z})^\times$.
Leander has pointed out~\cite{Le02} that if $b$
 is selected so that the Jacobi symbol $(b/N)=-1$ (``Choice L''), then, 
not only is the order $r$ of $b$ guaranteed to be even, but also the probability that $f_b=\gcd(b^{r/2}-1,N)$
is a non-trivial factor of $N$ is enhanced; for $r$ even, $f_b$ is the desired factor
provided $b^{r/2}\not\equiv-1(\Mod{N})$: the conditional probability
\begin{align}\label{eq:Leander}
 \Pr&\left( b^{r/2}\not\equiv-1(\Mod{N}) \;\middle\vert\; (b/N)=-1\right) \nonumber\\
         &\hspace*{50pt} = 1 - \frac{1}{2^{1+c_p  -c_q}}(1-\delta_{c_p,c_q}) ,
\end{align}
where the positive integer powers $c_p\ge c_q$ are related to the square-free semiprime $N$ by
the parametrisation $N=(2^{c_p} d_p + 1)(2^{c_q} d_q + 1)$,
it being understood that $d_p$ and $d_q$ are, by definition, odd. 
The success rate of the prescription for $f_b$ is at least 75\% as compared with 50\% if $b$ is chosen at
random from $(\mathbb{Z}/N\mathbb{Z})^\times$, but this
observation does not do justice to the full implications of \eqref{eq:Leander}.

The crux to building on Leander's criterion for $b$ is to recognise that, in principle, 
\emph{it can only fail to generate a factor of $N$ if $c_p>c_q$}. When it is known that $c_p > c_q$, 
then $b$ should be a quadratic non-residue modulo $N$ for which 
$(b/N)=+1$ (``Choice $\overline{\mathrm{L}}$''); the corresponding probability that $f_b$ is a prime 
factor of $N$ is $1-\delta_{c_p,c_q}$: if $c_p > c_q$, $f_b$ \emph{must\/} then be a non-trivial 
factor of $N$. (As for choice L, the fact that $b$ is 
a quadratic non-residue ensures that its order is even.)

There are two clear-cut complementary bench{\-}mark{\-}ing schemes $\mathcal{A}$ and $\mathcal{B}$: 
scheme $\mathcal{A}$ ($\mathcal{B}$) entails determination of the probability that $\overline{\mathrm{L}}$-choices 
(L-choices) for $b$ do actually yield a prime factor of $N$ given that $c_p>c_q$ ($c_p=c_q$).
Identification of whether $c_p>c_q$ or $c_p=c_q$ requires successful factorization of $N$, which can be 
achieved with a suitable sequence of initial runs,
beginning with a few L-choices, and converting to 
$\overline{\mathrm{L}}$-choices if these runs fail.
For both $\mathcal{A}$ and $\mathcal{B}$, 
the empirical probability is to be
compared with a predicted value of unity. The overhead placed on classical 
computer resources by the selection of appropriate values of $b$ is acceptable:
calculation of the Jacobi symbols is efficient, as is checking that a given $b\in  (\mathbb{Z}/N\mathbb{Z})^\times$ is a 
quadratic non-residue.

Application of the benchmarking procedures $\mathcal{A}$ and $\mathcal{B}$ should be preceded by determination of 
$(-1/N)$ and $(2/N)$. Provided $-1$ and 2 are quadratic non-residues modulo $N$, the values of these two Jacobi symbols
permit one to infer whether $c_p=c_q$ or $c_p>c_q$ for all semiprimes such that $c_q\le 2$ --- see Table \ref{tb:jsi}.
The import of Table \ref{tb:jsi} is that, for these semiprimes, the choice between $\mathcal{A}$ and $\mathcal{B}$ can be made
\emph{before\/} the order-finding algorithm is run. 
Blum integers are among the semiprimes to which the results of Table \ref{tb:jsi} apply.

\begin{table}[t]
\begin{onecolumn}
\centering
\begin{tabular}{ccc}
                     & {\small Interpretation}   &  Scheme  \\ \hline\hline
$(-1/N)=-1$   & $c_p>c_q=1$  & $\mathcal{A}$\\
$(-1/N)=+1$  & $c_p=1=c_q$  & $\mathcal{B}$\\
$(2/N)=-1$    & $c_p>c_q=2$  & $\mathcal{A}$\\
$(2/N)=+1$   & $c_p=2=c_q$  & $\mathcal{B}$ \\ \hline\hline
\end{tabular}
\caption{Choice of benchmarking scheme indicated by the Jacobi symbols $(-1/N)$ and $(2/N)$. 
The interpretation of $(-1/N)=+1$ and $(2/N)=+1$ assumes that $-1$ and 2, respectively, are
quadratic non-residues modulo $N$. If $-1\;(2)$ is a quadratic residue, then $c_q\ge 2\;(3)$.}
\label{tb:jsi}
\end{onecolumn}
\end{table}

It remains to justify the assertions made in this introduction. The exact reduction of the success probability 
associated with the period-finding algorithm to a form in \eqref{eq:genresult} suitable for controlled approximation
is presented in section \ref{sc:BW}, followed by its asymptotic analysis for large $r$ using the
Euler-Maclaurin summation formula in section \ref{sc:asymp}.
Properties of choices
L and $\overline{\mathrm{L}}$ are proven in section \ref{sc:Choices}, and some closing comments
are made in section \ref{sc:discuss}. Appendices \ref{app:A}, \ref{app:B} and \ref{app:C}
contain technical results required in sections  \ref{sc:BW} and \ref{sc:asymp}.

\section{Period-finding: success probability} \label{sc:BW}

After the usual preparatory steps (outlined, for example, in Algorithm 5 of Ref.~\cite{CvD10}), 
an $m$-qubit input register is left in the superposition of computational basis states
\begin{equation}\label{eq:superposition}
 \tfrac{1}{\sqrt{m_k}} \sum\limits_{l=0}^{m_l-1} |k+l\cdot r\rangle\hspace*{0.025\textwidth}
 \left( m_k = 1+ \left\lfloor \tfrac{2^m - 1 - k}{r} \right\rfloor \right),
\end{equation}
where $k\in\mathbb{Z}/r\mathbb{Z}$ is unknown (and unknowable), and the 
constraint on $m$ that
\begin{equation}\label{eq:constraint}
  2^m \ge r+ 1 + k > r 
\end{equation}
guarantees that the superposition contains more than one term and, 
hence, that $S_k(x)$ in \eqref{eq:StructureFactor} can manifest dependence on $r$. Since
\begin{equation}
r = r_\mathrm{o}2^{n_r}\hspace*{0.025\textwidth}
 \left(r_\mathrm{o}\ \mbox{odd} \right),
\end{equation}
where $n_r$ is a non-negative integer, an implication of \eqref{eq:constraint} found 
useful below is that $m>n_r$.

\begin{figure}[t]
  \centering
  \includegraphics[width=0.45\textwidth]{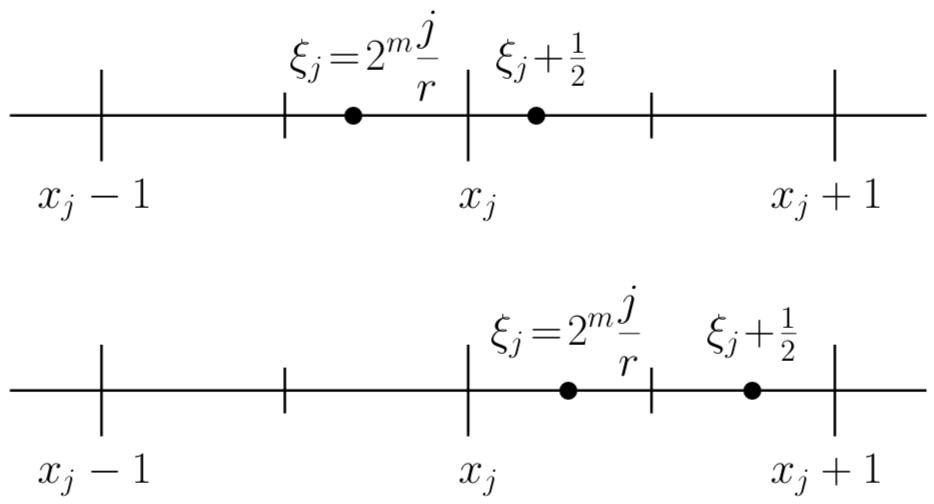}
  \caption{Confirmation that \eqref{eq:nearestinteger}, which is trivially valid if $2^m\tfrac{j}{r}$ is an integer, also
   works when $2^m\tfrac{j}{r}$ is a non-integer independent of whether $2^m\tfrac{ j}{r}<x_j$ or $2^m\tfrac{j}{r} >x_j$.}
  \label{fig:figure2}
\end{figure}

Interpretation of the quantum Fourier transform of the one-dimensional ``array'' of uniformly 
spaced ``atoms'' in \eqref{eq:superposition} is possible via its ``structure factor''
\begin{equation}\label{eq:StructureFactor}
 S_k(x)  =  \frac{1}{m_k}  \left| \sum\limits_{l=0}^{m_k-1} \left( e^{i 2\pi l}\right)^{r x/ 2^m} \right|^2 ,
\end{equation}
which is related to the conditional probability $P(x|k)$ that the transform is detected in the 
state $|x\rangle$ by $P(x|k)=S_k(x)/2^m$.
Central to the standard analysis of $P(x|k)$ is the observation that, for non-zero
integers $x$, $S_k(x)$ is large ($\sim m_k$) when the rational number $rx/2^m$, 
which must lie in the closed interval $[r/2^m,r(1-1/2^m)]$, is close to one of the $r-1$ 
integers $j$ in this interval, i.e.
\begin{equation}
 j\in\mathbb{N}_r^+=\{1,2,\ldots,r-1\} .
\end{equation}
Thus, the ``frequencies'' $x$ most likely to be returned by measurement are drawn from the
set of integers $\{x_j\}$ closest to the first $r-1$ members of the harmonic series with fundamental
``frequency'' $2^m/r$:
\begin{equation}\label{eq:LocusMaximum}
     \left| x_j - j  \frac{2^m}{r}\right| < \tfrac{1}{2}\hspace*{0.025\textwidth}
 \left( j \in \mathbb{N}^+_r \right),
\end{equation}
where the inequality is strict 
because $2^m\,j/r=2^{m-n_r}\, j/r_\mathrm{o}$ can never be a half-integer.
The solution of \eqref{eq:LocusMaximum} is
\begin{equation}\label{eq:nearestinteger}
   x_j  =\left\lfloor 2^m \tfrac{j}{r} +\tfrac{1}{2}\right\rfloor 
\end{equation}
as inspection of Fig.~\ref{fig:figure2} confirms.

Equation \eqref{eq:LocusMaximum} acquires additional significance if viewed from the perspective of rational
approximation with continued fractions. Provided the input register size is such that $2^{m/2}>r$, then
one of the finite number of convergents to $x_j/2^m$ coincides with the ratio $j/r$ reduced to lowest terms; 
if $m=\mathfrak{m}_0$, then the only observed values of $x$ from which information of this nature can be inferred
are precisely those belonging to the set $\{x_j\}$: the total probability for success of the period-finding algorithm is
\begin{equation}\label{eq:Ptotal}
 P_\mathrm{tot} = \sum\limits_{j\in\mathbb{N}^+_r} P(x_j|k) = \frac{1}{2^m} \sum\limits_{j\in\mathbb{N}^+_r} S_k(x_j) .
\end{equation}
More generally, when $m=\mathfrak{m}_\mathfrak{q}$, the restriction on useful values of $x$ reads
\begin{equation}\label{eq:usefulx}
    \left| x - j  \frac{2^{\mathfrak{m}_\mathfrak{q}}}{r}\right| < 2^{\mathfrak{q}-1} .
\end{equation}
When  $\mathfrak{q}\ge 1$, there are $2^\mathfrak{q}-1$ or $2^\mathfrak{q}$ solutions of \eqref{eq:usefulx}
dependent on whether $r_\mathrm{o}$ divides $j$ or not ($j\in \mathbb{N}_r^+$); 
if $\mathfrak{q}<0$, then the solutions are a subset of the $x_j$'s. 

\subsection{The case $\mathfrak{q}=0$}

Simplification of $P_\mathrm{tot}$ in \eqref{eq:Ptotal} for arbitrary $m \ge \mathfrak{m}_0$ prepares 
the ground for analysis of the general case when \eqref{eq:usefulx} applies.
The $j$-fold invocation of the periodicity property $S_k(x) = S_k(x+ 2^m/r)$ permits substitution of $S_k(x_j )$
by $S_k(\Delta_j )$, where
\begin{align}\label{eq:Delta}
  \Delta_j &= x_j - 2^m \frac{j}{r} \\
                                            & = \left\lfloor 2^{m-n_r} \tfrac{j}{r_\mathrm{o}} + \tfrac{1}{2}\right\rfloor 
                                                        - 2^{m - n_r}\tfrac{j}{r_\mathrm{o}}  ,  \nonumber
\end{align}
which, significantly, is periodic in $j$ with period $r_\mathrm{o}$. As a result, the partition of 
$\mathbb{N}^+_r$ into congruence classes modulo $r_\mathrm{o}$ serves to identify 
the summands $S(\Delta_j )$ for different $j$ in \eqref{eq:Ptotal} 
which are identical. For $r_\mathrm{o}>1$, the congruence classes 
$\overline{1}_{r_\mathrm{o}},\overline{2}_{r_\mathrm{o}} ,\ldots, \overline{r_\mathrm{o}-1}_{r_\mathrm{o}}$ in $\mathbb{N}^+_r$ 
all contain $r/r_\mathrm{o}(=2^{n_r})$ elements, but the congruence class $\overline{0}_{r_\mathrm{o}}$ has 
only $r/r_\mathrm{o}-1$ elements (because $0\not\in\mathbb{N}^+_r$); for $r_\mathrm{o}=1$, all 
$r-1\ (=r/r_\mathrm{o}-1)$ elements of $\mathbb{N}^+_r$ trivially belong to the single congruence 
class $\overline{0}_1$: thus,
\begin{equation}\label{eq:reducedPtotal}
P_\mathrm{tot} 
      = \frac{r}{r_\mathrm{o}} \sum\limits_{j\in\mathbb{Z}/r_\mathrm{o}\mathbb{Z}}  P(\Delta_j | k) -  P(\Delta_0|k)  ,
\end{equation}
where the sum over $j$ now includes $j=0$.

As shown in appendix \ref{app:A}, the properties of least non-negative residues modulo $r_\mathrm{o}$ imply that the $r_\mathrm{o}$ distinct 
values of $\Delta_j$ are equal to
$\mathfrak{j}/r_\mathrm{o}$, where the $\mathfrak{j}$'s are the absolute least residues
modulo $r_\mathrm{o}$:
\begin{equation}\label{eq:BZi}
  \mathfrak{j}\in\mathfrak{B}[r_\mathrm{o}] = \bigl\{0,\pm 1, \pm 2, \ldots , \pm\lfloor \tfrac{1}{2}r_\mathrm{o} \rfloor \bigr\} .
\end{equation}
The change of summation variable in \eqref{eq:reducedPtotal} from $j$ to $\mathfrak{j}=\mathfrak{j}(j)$
is indicated. The argument $\Delta_j$ is replaced by $\Delta_\mathfrak{j}= \mathfrak{j}/r_\mathrm{o}$
and
\begin{equation}\label{eq:Ptotq0}
 P_\mathrm{tot} 
  = \frac{r}{2^m} \left[\frac{1}{r_\mathrm{o}} \sum\limits_{\mathfrak{j}\in\mathfrak{B}[r_\mathrm{o}] } 
                                                S_k(\mathfrak{j}/r_\mathrm{o} ) - \frac{1}{r} S_k(0) \right] ,
\end{equation}
which suffices to analyse the case $m=\mathfrak{m}_0$.

\subsection{Generalization to $\mathfrak{q}<0$}

The result in \eqref{eq:Ptotq0} is easily
adapted to accommodate all cases in which $\mathfrak{q}<0$. Solutions of \eqref{eq:usefulx}
are those members of $\{x_j\}$ for which $|\Delta_j| < 1/2^{1-\mathfrak{q}}$. 
In terms of the index $\mathfrak{j}$ introduced in connection with \eqref{eq:BZi}, this inequality
specifies that $|\mathfrak{j}|<r_\mathrm{o}/2^{1-\mathfrak{q}}$ or, as $r_\mathrm{o}$ is odd,
\begin{equation}
   |\mathfrak{j}| \le \lfloor 2^{\mathfrak{q}-1}r_\mathrm{o} \rfloor .
\end{equation}
It follows that \eqref{eq:Ptotq0} is replaced by
\begin{equation}\label{eq:Ptotqneg}
 P_\mathrm{tot}\! 
  =\! \frac{r}{2^m}\! \left[\frac{1}{r_\mathrm{o}} \sum\limits_{\mathfrak{j}\in\mathfrak{B}[2^\mathfrak{q}r_\mathrm{o}] } 
                                                S_k(\mathfrak{j}/r_\mathrm{o} ) - \frac{1}{r} S_k(0) \right]\! .
\end{equation}
Unfortunately, the generalization to $\mathfrak{q}>0$ is not so immediate.

\subsection{Generalization to $\mathfrak{q}>0$}

If $j\in\overline{0}_{r_\mathrm{o}}$, then \eqref{eq:usefulx} amounts to the inequality
$
  \left| x - x_j  \right| < 2^{\mathfrak{q}-1},
$
which has the integer solutions $x_j  + \kappa$ with
\begin{equation}\label{eq:kappa0}
 \kappa \in \mathfrak{S}_0= \{0,\pm 1,\pm 2, \ldots,\pm(2^{\mathfrak{q}-1}-1) \}.
\end{equation}
For members of the remaining congruence classes summed over in
\eqref{eq:Ptotq0}, \eqref{eq:usefulx}
can be recast in terms of the congruence class label $\mathfrak{j}$ of \eqref{eq:BZi} as
$
  \left| x - x_j + \mathfrak{j}(j)/r_\mathrm{o}  \right| < 2^{\mathfrak{q}-1} .
$
Now, the set of integer solutions $\{x_j+\kappa\}$ depends on $\mathfrak{j}=\mathfrak{j}(j)$ via its sign:
$\kappa$ takes on all values in the set
\begin{equation}\label{eq:kappanonzero}
\mathfrak{S}_\mathfrak{j}=\mathfrak{S}_0\cup\{-(\mathrm{sgn}\,\mathfrak{j}) 2^{\mathfrak{q}-1} \} ,
\end{equation}
where $\mathrm{sgn}\,\mathfrak{j}$ denotes the sign of the non-zero $\mathfrak{j}$ in \eqref{eq:kappanonzero}.

Ostensibly, the summation in Eq, \eqref{eq:Ptotq0} is replaced by
the double summation
\begin{equation}
 \sum\limits_{\mathfrak{j}\in\mathfrak{B}[r_\mathrm{o}] } \,
 \sum\limits_{\kappa\in\mathfrak{S}_{\mathrm{j}} }
                                                S_k(\kappa+\mathfrak{j}/r_\mathrm{o} ) ,
\end{equation}
but it can be more usefully rewritten in terms of a single summation formally like that in \eqref{eq:Ptotqneg} as
\begin{equation}\label{eq:singlesum}
 \sum\limits_{\mathfrak{j}\in\mathfrak{B}[2^\mathfrak{q}r_\mathrm{o}]}\!\! S_k(\mathfrak{j}/r_\mathrm{o})\,
 - \epsilon_k ( \mathfrak{q} ) ,
\end{equation}
provided the endpoint ``correction''
\begin{equation}
\epsilon_k ( \mathfrak{q} ) = S_k(2^{\mathfrak{q}-1}) + S_k(-2^{\mathfrak{q}-1}) 
\end{equation}
is included.
A simpler substitution is that of the coefficient $S_k(0)$ of $1/r$ in \eqref{eq:Ptotq0} by
\begin{equation}\label{eq:coefficientsum}
  \sum\limits_{\kappa\in\mathfrak{S}_0 } S_k(\kappa) = 
   \sum\limits_{\kappa\in\mathfrak{B}[ 2^\mathfrak{q} ]}\!\! S_k(\kappa )\,
 - \epsilon_k ( \mathfrak{q} ) .
\end{equation}
Together, \eqref{eq:singlesum} and \eqref{eq:coefficientsum} imply that the generalisation 
to positive $\mathfrak{q}$ of \eqref{eq:Ptotq0} is
\begin{align}\label{eq:genresult}
 P_\mathrm{tot}^{(\mathfrak{q})} 
  = \frac{r}{2^{\mathfrak{m}_\mathfrak{q}}} & \Biggl[\frac{1}{r_\mathrm{o}} 
             \sum\limits_{\mathfrak{j}\in\mathfrak{B}[2^{\mathfrak{q}}r_\mathrm{o}] } \!\!
       S_k(\mathfrak{j} / r_\mathrm{o} ) - \frac{1}{r} \sum\limits_{\mathfrak{j}\in\mathfrak{B}[ 2^\mathfrak{q} ]} S_k(\mathfrak{j}) 
                                                \nonumber \\[1ex]
        & \hspace*{0.1\textwidth}   - \left(\frac{1}{r_\mathrm{o}} -  \frac{1}{r} \right) \epsilon_k ( \mathfrak{q} )     \Biggr]  .                                                            
\end{align}
The choice of $2^m$ appropriate to \eqref{eq:genresult} (namely, $2^{\mathfrak{m}_\mathfrak{q}}$) 
has been made for the
overall multiplicative factor; it has also to be adopted in the expression for $S_k(x)$ 
in \eqref{eq:StructureFactor}.
With the extended definition of $\epsilon_k (\mathfrak{q})$ in terms of the Heaviside function $H(x)$ 
(\cite{DLMF}, Eq.~1.16.13) as
\begin{equation}
 \epsilon_k (\mathfrak{q}) = H(\mathfrak{q}) \left[ S_k(2^{\mathfrak{q}-1}) + S_k(-2^{\mathfrak{q}-1}) \right] ,
\end{equation}
 \eqref{eq:genresult} continues to hold when $\mathfrak{q}\le 0$.

Equation \eqref{eq:genresult} is the goal of this section.
It is a generalization and refinement of results in \cite{BW07}, which, in addition to the features 
mentioned in the introduction, includes a more careful treatment of endpoint corrections. 
The derivation has invoked only the periodicity of
$S_k(x)$ and is independent of its precise form, which, in \eqref{eq:genresult}, happens to be
\begin{equation}\label{eq:explicitS}
 S_k(x) = S_k(0) 
      \left[ \frac{\sinc(m^{(\mathfrak{q})}_k  r x /2^{\mathfrak{m}_\mathfrak{q}})}{ \sinc(
                                r x / 2^{\mathfrak{m}_\mathfrak{q}} )} \right]^2 ,
\end{equation}
where $S_k(0)=m_k^{(\mathfrak{q})} = 1+ \left\lfloor (2^{\mathfrak{m}_\mathfrak{q}} - 1 - k)/r \right\rfloor $
and $\sinc(x)$ is 
the \emph{normalized\/} \mbox{sinc} function [for $x\not=0$, $\sinc(x)=\sin(\pi x)/(\pi x)$].

The endpoint correction term in \eqref{eq:genresult} does not influence the approximate 
estimates of $P_\mathrm{tot}^{(\mathfrak{q})}$ considered next. Terms involving non-zero
$\epsilon_k(\mathfrak{q})$ are suppressed by at least three powers of $r$ relative to the dominate large $r$
contribution.

\section{Asymptotic analysis of $P_\mathrm{tot}^{(\mathfrak{q})}$}   \label{sc:asymp}

If $r$ is a power of two (i.e.\, $r_\mathrm{o}=1$), then $m_k^{(\mathfrak{q})} = 2^{\mathfrak{m}_\mathfrak{q}} / r$,
$S_k(x)=\delta_{x,0}\, m_k^{(\mathfrak{q})}$ (for integer $x$), and \eqref{eq:genresult} reduces without approximation 
to
\begin{equation}\label{eq:ro1}
  P_\mathrm{tot}^{(\mathfrak{q})}  = 1 - \frac{1}{r} .
\end{equation}
The first summation in \eqref{eq:genresult} is dominant, and the remaining terms are suppressed by one power of $r$
or more. The same pattern is also evident in an expansion of  \eqref{eq:genresult} for large $r$ which embraces 
the cases in which $r$ is \emph{not\/} a power of two. If the negligible terms of order $1/r$ and higher
are discarded, then the following approximation to $P_\mathrm{tot}^{(\mathfrak{q})}$ emerges:
\begin{equation}\label{eq:domC}
   \mathsf{P}^{(\mathfrak{q})}
  = \frac{1}{r_\mathrm{o}} 
             \sum\limits_{\mathfrak{j}\in\mathfrak{B}[2^{\mathfrak{q}}r_\mathrm{o}] } \!\!
       \sinc^2 (\mathfrak{j} / r_\mathrm{o} ) .
\end{equation}
The expansions in support of \eqref{eq:domC} are given in appendix \ref{app:B}. Like  \eqref{eq:ro1},
\eqref{eq:domC} is independent of $k$, but, when specialized to $r_\mathrm{o}=1$, yields 
$\mathsf{P}^{(\mathfrak{q})}=1$.

\begin{table}[t]
\centering
\caption{Sample of dependence of $\mathsf{P}^{(\mathfrak{q})}$ on $r_\mathrm{o}$ for $\mathfrak{q}\ge0$. 
The last row contains
the values of  $\mathsf{P}^{(\mathfrak{q})}_\infty$ from \eqref{eq:Pq}. }\label{tb:qc2}
\begin{tabular}{llll} \hline 
 $r_\mathrm{o}\backslash \mathfrak{q} $  &  0           &  1                 &  2                     \\ \hline
                                                   3   &  0.7893  &  0.90326      &  0.949999  \\
                                                   5   &  0.7792  &  0.90288      &  0.949946  \\
                                                   7   &  0.7765  &  0.902837    &  0.9499411  \\  
                                                   9   &  0.7754  &  0.902828    &  0.9499400  \\
                                                 11   &  0.7748  &  0.902826    &  0.9499396  \\
                                                 13   &  0.7745  &  0.9028245  &  0.94993949  \\
                                                 15   &  0.7743  &  0.9028240  &  0.94993942  \\ \hline
                                          $\infty$  &  0.7737  &   0.9028233  & 0.94993934  \\ \hline
\end{tabular}
\end{table}

Simple as the expression for $\mathsf{P}^{(\mathfrak{q})}$ is, it would seem of little utility because of its
reliance on the factor $r_\mathrm{o}$ of the unknown (but large) period $r$. 
However, the numerical data in Table \ref{tb:qc2} suggests that 
$\mathsf{P}^{(\mathfrak{q})}$ is a monotonically decreasing function of $r_\mathrm{o}$ when $\mathfrak{q}\ge0$.
This trend can be confirmed for large $r_\mathrm{o}$ with the aid of the Euler-Maclaurin 
summation formula, which, for $\mathfrak{q}\ge1$, implies straightforwardly that
\begin{equation}\label{eq:asympq}
 \mathsf{P}^{(\mathfrak{q})} = \frac{2}{\pi}\mathrm{Si}(2^\mathfrak{q}\pi) +
                                                     \frac{4}{15\cdot 8^\mathfrak{q}}\frac{1}{r_\mathrm{o}^4} + \ldots\, .
\end{equation}
The calculation is trickier for $\mathfrak{q}=0$; application of the Euler-Maclaurin summation formula
has to be followed by an expansion in inverse powers of $r_\mathrm{o}$ (details are given in
appendix \ref{app:C}): the outcome of these manipulations is
\begin{equation}\label{eq:asymp0}
 \mathsf{P}^{(\mathfrak{q}=0)} =  \frac{2}{\pi}\left(\mathrm{Si}(\pi) -\frac{2}{\pi} \right)  
                                                     +  \frac{4}{3 \pi^2 }\frac{1}{r_\mathrm{o}^2} + \ldots\, . 
\end{equation}
The results in \eqref{eq:asympq}, \eqref{eq:asymp0} and Table \ref{tb:qc2} 
justify the claim that, \emph{provided $1/r$ is negligible, $\mathsf{P}^{(\mathfrak{q})}_\infty$ in \eqref{eq:Pq} is always
 an excellent approximation to $P_\mathrm{tot}^{(\mathfrak{q})}$} in \eqref{eq:genresult} when $\mathfrak{q}$ is non-negative.

\begin{table}[t]
\setlength{\tabcolsep}{4pt}
\centering
\caption{Sample of deviation of $\mathsf{P}^{(\mathfrak{q})}$ from $\mathsf{P}^{(\mathfrak{q})}_\infty$ for $\mathfrak{q}<0$. 
The scaled deviation $\mathsf{D}^{(\mathfrak{q})}=r_\mathrm{o}\bigl(\mathsf{P}^{(\mathfrak{q})} - \mathsf{P}^{(\mathfrak{q})}_\infty\bigr)$. 
The last row contains the values of $\mathsf{D}^{(-2)}$ in the limit $r_\mathrm{o}\rightarrow\infty$.}\label{tb:qc3}
\begin{small}
\begin{tabular}{clclclcl} \hline 
 $r_\mathrm{o}$  &  $\mathsf{D}^{(-2)}$  &  $r_\mathrm{o}$  &  $\mathsf{D}^{(-2)}$  & $r_\mathrm{o}$  &  $\mathsf{D}^{(-2)}$  & $r_\mathrm{o}$  &  $\mathsf{D}^{(-2)}$ \\ \hline
                            &                                  &         3                  &  0.263                       &        5                  &  $-$0.229                   &        7                  &  $-$0.720                 \\
        9                  &  0.708                       &         11                &  0.243                       &        13                &  $-$0.234                    &       15                &  $-$0.716                 \\
        17                &  0.7098                     &         19                &  0.240                       &        21                &  $-$0.235                    &       23                &  $-$0.7144               \\
        25                &  0.7105                     &         27                &  0.239                       &        29                &  $-$0.236                    &       31                &  $-$0.7138              \\ \hline
                           & 0.7122                       &                             &  0.237                       &                            &  $-$0.237                    &                           &  $-$0.7122              \\  \hline 
\end{tabular}
\end{small}
\end{table}

\begin{figure}[b!] 
  \centering
  \includegraphics[width=0.45\textwidth]{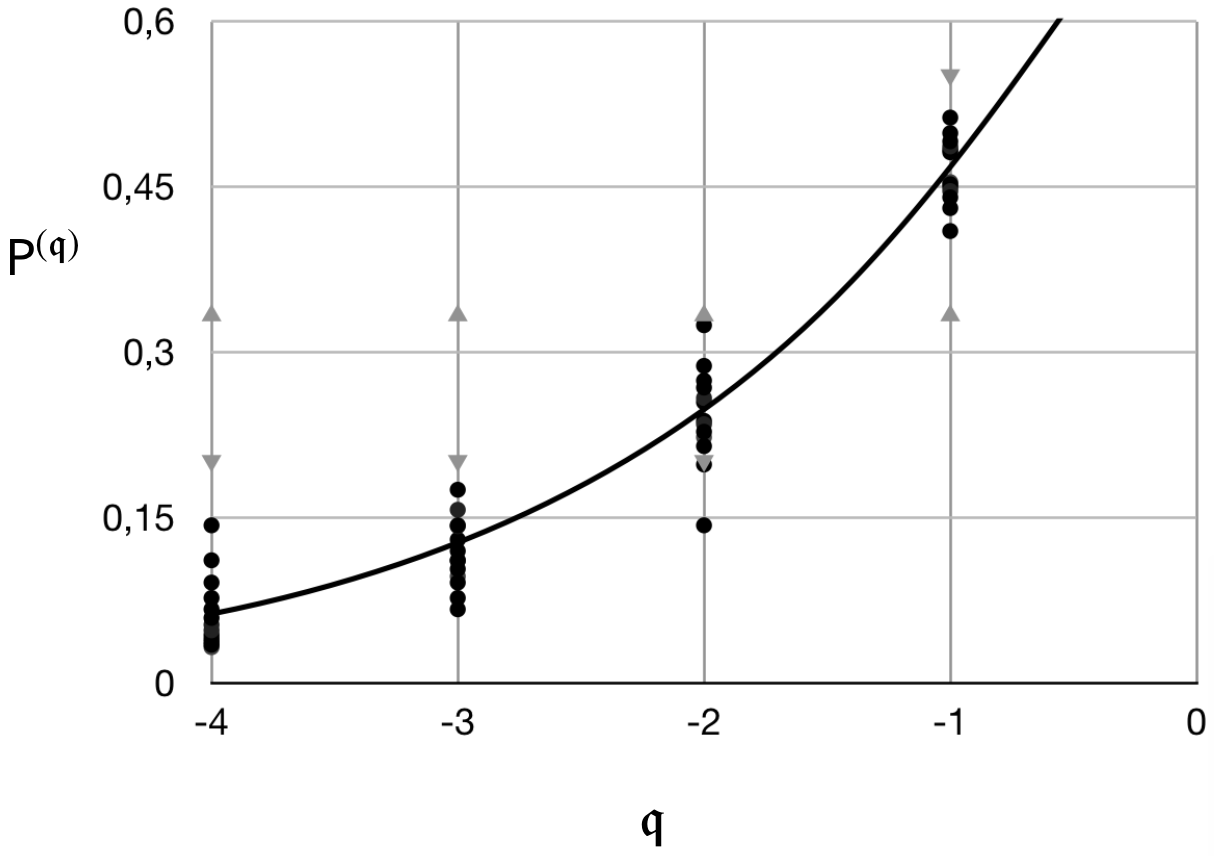}
  \caption{Comparison of values of $\mathsf{P}^{(\mathfrak{q})}$ in \eqref{eq:domC} to $\mathsf{P}_\infty^{(\mathfrak{q})}$ 
  for \emph{negative\/} increments $\mathfrak{q}$ of the input register size from its critical size (of $\mathfrak{m}_0$ qubits). 
  Values for which $r_\mathrm{o}=3\, (5)$ are represented by the symbol
  {\color{lightgray}$\blacktriangle\,$}({\color{lightgray}$\blacktriangledown$}); the symbol $\bullet$ is used for all larger values of
  $r_\mathrm{o}$ considered ($7\le r_\mathrm{o}\le 31$).
}  \label{fig:figure3}
\end{figure}

The behaviour of $\mathsf{P}^{(\mathfrak{q})}$ for $\mathfrak{q}<0$ is quite different. Now, the expansion in inverse
powers of $r_\mathrm{o}$, which is obtained along the same lines as \eqref{eq:asymp0}, reads
\begin{align}
  \frac{2}{\pi}\mathrm{Si}(2^\mathfrak{q} \pi) & - 2^\mathfrak{q} \sinc^2( 2^{\mathfrak{q}-1}) \label{eq:asympn} \\
  & + \sinc^2 (2^{\mathfrak{q}-1})\left( 1- \frac{\nu}{2^{|\mathfrak{q}|}} \right) \frac{1}{r_\mathrm{o}} + \ldots , \nonumber
\end{align}
where $\nu$ is the least non-negative residue of $r_\mathrm{o}$ modulo $2^{|\mathfrak{q}|+1}$. The correction to 
the leading term $\mathsf{P}^{(\mathfrak{q})}_\infty$ is of either sign and,  being of order $1/r_\mathrm{o}$
(see Table \ref{tb:qc3}),
can be substantial for small $r_\mathrm{o}$ -- see Fig.~\ref{fig:figure3}. Despite the scatter  about 
$\mathsf{P}^{(\mathfrak{q})}_\infty$ in the values of $\mathsf{P}^{(\mathfrak{q})}$ for $\mathfrak{q}<0$, 
none are close to $\mathsf{P}^{(0)}$.

The leading term 
$\mathsf{P}^{(\mathfrak{q})}_\infty$ in all of the above expansions in \eqref{eq:asympq}, \eqref{eq:asymp0} and
\eqref{eq:asympn} is found by evaluation of 
\begin{equation}\label{eq:intrep}
  \int\limits_{-2^{\mathfrak{q}-1}}^{2^{\mathfrak{q}-1}} \sinc^2(x) dx  
\end{equation}
for integer $\mathfrak{q}$. The integral in \eqref{eq:intrep} defines an entire function of $z=2^\mathfrak{q}$. 
In the limit $r_\mathrm{o}\rightarrow\infty$, when $\mathsf{P}^{(\mathfrak{q})}\rightarrow\mathsf{P}^{(\mathfrak{q})}_\infty$,
the lower limit on  the integers $\mathfrak{q}$ of $\mathfrak{q}_{\min}\rightarrow-\infty$; the values of $2^\mathfrak{q}$ 
then belong to a set with an accumulation point, and the identity theorem for holomorphic functions
can be invoked to assert that the integral representation in \eqref{eq:intrep} is the unique
analytic continuation of $\mathsf{P}^{(\mathfrak{q})}_\infty$ to all values of $\mathfrak{q}$, real and complex: thus,
the right-hand side of \eqref{eq:Pq}, which is obtained from \eqref{eq:intrep} for arbitrary 
$\mathfrak{q}$ by integration by parts, is a natural choice of interpolation function in Fig.~\ref{fig:figure1}.

\section{Foundations for choices L and $\overline{\mathrm{L}}$} \label{sc:Choices}

Insight into good choices of $b$ for the factorization of square-free odd semiprimes
$N=pq$ is gained through the isomorphism 
between $(\mathbb{Z}/N\mathbb{Z})^\times$
and the direct product $(\mathbb{Z}/p\mathbb{Z})^\times \times (\mathbb{Z}/q\mathbb{Z})^\times$
of the cyclic groups $(\mathbb{Z}/p\mathbb{Z})^\times$ and $(\mathbb{Z}/q\mathbb{Z})^\times$.
The Jacobi symbol permits some consequences of this isomorphism
to be recast in a manner which does not require any knowledge of the odd primes $p$ and $q$.

\subsection{Use of $(\mathbb{Z}/N\mathbb{Z})^\times\cong(\mathbb{Z}/p\mathbb{Z})^\times
 \times (\mathbb{Z}/q\mathbb{Z})^\times$ } \label{sbsc:isomorphism}

The two simultaneous congruence relations
\begin{equation}\label{eq:bcong}
 b\equiv b_p(\Mod{p}),\quad b\equiv b_q(\Mod{q})
\end{equation}
establish (via the Chinese Remainder theorem) a bijection between elements 
$b\in (\mathbb{Z}/N\mathbb{Z})^\times$ and ordered pairs 
$(b_p,b_q)\in (\mathbb{Z}/p\mathbb{Z})^\times \times (\mathbb{Z}/q\mathbb{Z})^\times$, 
and there is a one-to-one correspondence between $b^k\,(\Mod{N})$ and $\bigl(b_p^k(\Mod{p}),
b_q^k(\Mod{q})\bigr)$ for any positive integer power $k$. Hence, the order $r$ of $b$ to be 
used in $f_b=\gcd(b^{r/2}-1,N)$ is determined by the orders $r_p$  and $r_q$ of $b_p$ and 
$b_q$, respectively, via
\begin{equation}\label{eq:rlcm}
   r = \lcm (r_p,r_q) ,
\end{equation}
and the \emph{conditions for the failure of $f_b$ as a factor of $N$\/}
find neat expression as a property of $r_p$ and $r_q$:
\emph{the powers of two in the prime factorizations of $r_p$ and $r_q$ are equal\/}  
(\cite{CvD10}, Lemma 2).

The appeal of this characterisation is that the distribution of the powers of two associated with 
either of the orders $r_p$ or $r_q$ is easily constructed. For odd primes $h=2^c d+1$ 
($c\ge 1$, $d$ odd), the group $(\mathbb{Z}/h\mathbb{Z})^\times$ comprises elements which 
are powers (modulo $h$) of a generator $\mathfrak{g}_h$ of order $h-1=2^c d$. As a result, the element 
$\mathfrak{g}_h^k(\Mod{h})$ (where $k=1,2,\ldots,2^c d$) has order
\begin{equation}\label{eq:orderGENERAL}
  r_h^{(k)} = \frac{2^c d}{\gcd(k,2^cd)} .
\end{equation}
For the $2^{c-1} d$ \textit{odd\/} values of $k$, 
\eqref{eq:orderGENERAL} simplifies to $r_h^{(k)} = 2^c d_k$,
where the odd number 
$d_k = d/\gcd(k,d)$; the corresponding result for the equal number of
 \textit{even\/} values of $k=2j$ ($ j = 1,2,\ldots,2^{c-1}d$) is
\begin{equation}\label{eq:orderEVEN}
   r_h^{(k=2j)} = \frac{2^{c-1} d}{\gcd(j,2^{c-1}d)}   ,
\end{equation}
which resembles \eqref{eq:orderGENERAL} with the factor of $2^c$ replaced by $2^{c-1}$:
on the basis of the recursive pattern implied by the similarity of \eqref{eq:orderEVEN} to \eqref{eq:orderGENERAL}, there are
$2^{l-1}d$ elements of $(\mathbb{Z}/h\mathbb{Z})^\times$ with the \textit{even\/} orders
$2^l d_j$ ($ j=1,3,5,\ldots, 2^l d-1$) for each $l\in\{1,2,3,\ldots,c\}$, and only $d$ elements 
with the \textit{odd\/} orders $d_j$ ($j=1,2,\ldots,d$).
Altogether, there are $c+1$
different powers of two: $\{0,1,\dots, c-1\}$ for the even choices of $k$ in 
\eqref{eq:orderGENERAL}, and exclusively $c$ for the odd choices. 
 
An element $(b_p,b_q)\in(\mathbb{Z}/p\mathbb{Z})^\times\times (\mathbb{Z}/q\mathbb{Z})^\times$ has
the representation $\bigl(\mathfrak{g}_p^{k_p}(\Mod{p}),\mathfrak{g}_q^{k_q}(\Mod{q})\bigr)$ in terms 
of generators $\mathfrak{g}_p$ and $\mathfrak{g}_q$ of $(\mathbb{Z}/p\mathbb{Z})^\times$ and 
$(\mathbb{Z}/q\mathbb{Z})^\times$, respectively. Consistent with the earlier parametrization of $N$ in 
connection with \eqref{eq:Leander}, the primes $p$ and $q$ are taken to be $p=2^{c_p}d_p+1$ and 
$q=2^{c_q}d_q+1$, where $d_p, d_q$ are odd and $c_p\ge c_q$. 
The results of the previous paragraph can be used to determine the number of pairs $(b_p,b_q)$
for which the powers of two in the prime factorizations of the corresponding orders $r_p$ and $r_q$ 
are identical. By way of example, whenever both indices $k_p$ and $k_q$ are \emph{odd}, the respective 
powers of two are $c_p$ and $c_q$; as there are $\tfrac{1}{2}(p-1)$ odd values of $k_p$ and  
$\tfrac{1}{2}(q-1)$ odd values of $k_q$, the corresponding
number of pairs $(b_p,b_q)$ with orders sharing the same power of two is
\begin{equation}
\tfrac{1}{2}(p-1){\times} \tfrac{1}{2}(q-1)\hspace*{0.5pt}\delta_{c_p,c_q} 
                                                                = 4^{c_q - 1} d_p d_q\hspace*{0.5pt} \delta_{c_p,c_q} .
\end{equation}
Table \ref{tb:qc4} contains a summary of all the findings on the numbers of pairs with such matching powers of two.
No match is possible when $k_p$ is \emph{odd\/} and $k_q$ is \emph{even\/} because $c_q-1<c_p$.
The entry for \emph{even} $k_p$ and $k_q$ excludes the pair $(1,1)$ since it corresponds to a
choice of $b$ ($b=1$) that cannot yield non-trivial factors of $N$ ($f_{b=1}=N$).

\begin{table}[t]\centering
\caption{The numbers of pairs $(b_p,b_q)\in
(\mathbb{Z}/p\mathbb{Z})^\times\times (\mathbb{Z}/q\mathbb{Z})^\times$
for which the powers of two in the prime factorizations of the two related orders $r_p$ and $r_q$
coincide. The sum $S=4^{c_q-1}d_pd_q$. \\[-1em]  }\label{tb:qc4} 
\begin{small}
\begin{tabular}{|c|cc|} \cline{2-3} 
  \multicolumn{1}{c|}{}        & \rule[-6pt]{0pt}{18pt}   $k_q$ \small{even}   &    $k_q$ \small{odd}                   \\ \hline
\rule[-6pt]{0pt}{18pt}  
\strut $k_p$ \small{even}  &  $\tfrac{1}{3}S+\tfrac{2}{3} d_pd_q-1$            &  $S(1-\delta_{c_p,c_q})$ \\
\rule[-6pt]{0pt}{18pt}
$k_p$ \small{odd}             &   0                                                                &  $S\hspace*{0.5pt}\delta_{c_p,c_q}$  \\ \hline
\end{tabular}
\end{small}
\end{table}

\subsection{Role of the Jacobi symbol}
 
Quadratic residues modulo $N$ are those elements of $(\mathbb{Z}/N\mathbb{Z})^\times$ for which
both of the indices $k_p$ and $k_q$ are even. 
Table \ref{tb:qc4} exhibits a partition of
$(\mathbb{Z}/N\mathbb{Z})^\times$ into the subgroup comprising
its quadratic residues and the three cosets of this subgroup, all elements of which are quadratic non-residues modulo $N$. 
The Jacobi symbol distinguishes two of these cosets from the subgroup of quadratic residues.

For odd primes $h$, it is the Legendre symbol $(a/h)$ which differentiates between quadratic residues and 
non-residues modulo $h$: $(a/h)$ is $+1$ ($-1$) if $a$ is a quadratic residue (non-residue), and 0 if
$h$ is a divisor of $a$. 
A consequence of Fer{\-}mat's little theorem is that $a^{(h-1)/2}\,(\Mod{h})$
must have one of $+1$, $-1$ or 0 as its least absolute residue. As a result (\cite{HW08}, Theorem 83),
it is possible to express $(a/h)$ in terms of least absolute residues as
\begin{equation}
 \left(\frac{a}{h}\right) \equiv a^{(h-1)/2}\,(\Mod{h})  .
\end{equation}
Furthermore, as
the values $+1$ and 0 are inadmissible for any 
generator $\mathfrak{g}_h$ of $(\mathbb{Z}/h\mathbb{Z})^\times$
because they are incompatible with the requirement that it be of even order $h-1$, it is necessarily the case
that $\mathfrak{g}_h^{(h-1)/2}\equiv-1 (\Mod{h})$, and
\begin{align}\label{eq:interLC}
 \left(\frac{\mathfrak{g}_h^k}{h}\right)  &\equiv \bigl(\mathfrak{g}_h^{(h-1)/2}\bigr)^k\,(\Mod{h}) \\
                                                            &\equiv (-1)^k\, (\Mod{h})  = (-1)^k  \nonumber  ,
\end{align}
in accord with the expectation that even (odd) powers of a generator are quadratic residues (non-residues).

For an integer $b$ relatively prime to the square-free semiprime $N=pq$, the Jacobi symbol
\begin{equation}\label{eq:Jdef}
 \left(\frac{b}{N}\right) = \left(\frac{b}{p}\right) \left(\frac{b}{q}\right),
\end{equation}
where the right-hand side is the product of the two Legendre symbols $(b/p)$ and $(b/q)$, 
which, in view of the congruences in \eqref{eq:bcong}, can be substituted by $(b_p/p)$ 
and $(b_q/q)$, respectively. Use of the further congruences
$b_p\equiv(\mathfrak{g}_p)^{k_p}\,(\Mod{p})$ and $b_q\equiv (\mathfrak{g}_q)^{k_q}\,(\Mod{q})$
as well as \eqref{eq:interLC} imply finally that
\begin{equation}
  \left(\frac{b}{N}\right) = (-1)^{k_p+k_q} ,
\end{equation}
which is the basis for the observation exploited in \cite{Le02} that, when $(b/N)=-1$, $b$ 
belongs to the union of the ($k_p$ odd, $k_q$ even) and 
($k_p$ even, $k_q$ odd) cosets in $(\mathbb{Z}/h\mathbb{Z})^\times$.

According to Table \ref{tb:qc4}, $S(1-\delta_{c_p,c_q})$ of the $\tfrac{1}{2}(p-1){\times} (q-1) = 2^{c_p-c_q+1}S$
members in this union are unsuitable for the purposes of factoring the semiprime $N$. The result for the
conditional probability in \eqref{eq:Leander} follows immediately. Another inference from Table  \ref{tb:qc4},
which forms the basis for choice $\overline{\mathrm{L}}$, is that, when it is known $c_p\not=c_q$, then
\emph{all\/} quadratic non-residues $b$ for which $(b/N)=+1$ [i.e. the whole of the ($k_p$ odd, $k_q$ odd) coset] 
are good candidates for factoring $N$.

\subsection{Value of $c_q$: special cases}

Specialization 
to the square-free semiprime $N=(2^{c_p}d_p+1)(2^{c_q}d_q+1)$ of
the standard formulae for the Jacobi symbols $(-1/N)$ and $(2/N)$ (\cite{DLMF}, \S 27.9)
proves serendipitously fruitful.
To begin with, on the basis of
\begin{align}\label{eq:minus1}
\left(\frac{-1}{N}\right)  &=  (-1)^{(N-1)/2}        \\
                                    &=  (-1)^{2^{c_p-1}}  (-1)^{2^{c_q-1}} , \nonumber
\end{align}
it is possible to interpret the value of $(-1/N)$ as follows:
if $(-1/N)=-1$, then, without further ado, $c_p>c_q=1$; if, instead, $(-1/N)=+1$, then
$c_p=1=c_q$ when $-1$ is a quadratic \emph{non\/}-residue modulo $N$, 
and otherwise it can be deduced that $c_p\ge c_q\ge 2$.
In the latter case, it is appropriate to move onto $(2/N)$. Under the assumption
that 
$c_p\ge c_q\ge 2$,
\begin{align}\label{eq:plus2}
 \left(\frac{2}{N}\right) &=  (-1)^{(N^2-1)/8}        \\
                                   &=  (-1)^{2^{c_p-2}}  (-1)^{2^{c_q-2}} , \nonumber
\end{align}
the implications of which parallel those of \eqref{eq:minus1}: if $(2/N)=-1$, then
$c_p>c_q=2$; if $(2/N)=+1$, then $c_p=2=c_q$ when $+2$ is a quadratic 
\emph{non\/}-residue modulo $N$, and, by the exclusion above of other options, 
$c_p\ge c_q\ge 3$ when $+2$ is a quadratic residue. All of these findings are summarized in
Table \ref{tb:jsi}.

Larger values of $c_q$ can be identified if it is known that $c_p>c_q$ 
(because choice L has failed) by the elementary expedient of evaluating
\begin{equation}
  s_k = (-1)^{(N-1)/2^k} 
\end{equation}
for $k=3,4,\ldots$. As $s_k = (-1)^{2^{c_q-k}}$ for $k\le c_q<c_p$, the sequence of
evaluations is to be terminated when the value $s_k=-1$ is encountered; $c_q$ is the 
corresponding value of $k$.

\subsection{Properties of orders}

With the substitutions $r_p=2^{l_p}r_{p\mathrm{o}}$ and $r_q=2^{l_q}r_{q\mathrm{o}}$ 
($r_{p\mathrm{o}}, r_{q\mathrm{o}}$ odd), \eqref{eq:rlcm} becomes
\begin{equation}
 r = 2^{{\max}(l_p,l_q)}\lcm(r_{p\mathrm{o}},r_{q\mathrm{o}}) .
\end{equation}
The properties of the indices $l_p$ and $l_q$ established in subsection \ref{sbsc:isomorphism} imply that,
for choice L,
\begin{equation}
c_q  \le  \max(l_p,l_q)\le c_p,
\end{equation}
whereas, for choice $\overline{\mathrm{L}}$,
\begin{equation}
  \max(l_p,l_q) =  c_p.
\end{equation}
For both choices, the order $r$ is even as asserted in the introduction.

Lagrange's theorem for finite groups and the isomorphism 
$(\mathbb{Z}/N\mathbb{Z})^\times\cong(\mathbb{Z}/p\mathbb{Z})^\times
\times (\mathbb{Z}/q\mathbb{Z})^\times$ imply that an order $r$ modulo the
square-free semiprime $N$ 
is a divisor of the value
\begin{equation}
 \lambda(N) = \lcm(p-1,q-1)
\end{equation}
of the Carmichael $\lambda$-function. 
Thus, substituting for $\lcm(p-1,q-1)$
in terms of $\gcd(p-1,q-1)$,
\begin{equation}\label{eq:rbound}
 r \le \frac{(p-1)(q-1)}{\gcd(p-1,q-1)}\le\frac{1}{2^{c_q}} (p-1)(q-1)
\end{equation}
since $\gcd(p-1,q-1)\ge 2^{c_q}$. 
In all cases of practical interest, $p+q>2$ and
the right-hand side of \eqref{eq:rbound} can be replaced by
\begin{equation}\label{eq:rmax}
 r_{\max} = \tfrac{1}{2^{c_q}} (pq-1) =\tfrac{1}{2^{c_q}} (N-1) .
\end{equation}
Information gleaned from the analysis of the Jacobi symbols $(-1/N)$ and $(2/N)$
or the ad hoc construct $s_k$
can be used to fix a suitable lower limit to $c_q$. The upper bound $r_{\max}$ can be used to
improve on Shor's recommendation that the input quantum 
register contain at least $m_{\mbox{\tiny Sh}}=\lceil 2\log_2N\rceil$ qubits.
In terms of $\mathfrak{m}_{\mathfrak{q}}$,
\begin{equation}
 m_{\mbox{\tiny Sh}}=\mathfrak{m}_{\mathfrak{q}=2c_q+\Delta} ,
\end{equation}
where $\Delta=\lceil 2\log_2 (N/2^{c_q})\rceil-\lceil 2\log_2r\rceil$ is a non-negative integer.

\section{Discussion} \label{sc:discuss}

As a tool for factoring RSA integers $N$, Shor's algorithm has been displaced by an approach 
which computes discrete logarithms~\cite{GE21,GS21}. Nevertheless, the present paper suggests that
there may remain an alternative use for Shor's algorithm as a context for testing the operation of quantum 
computers. The benchmarks involving quadratic non-residues (schemes $\mathcal{A}$ and $\mathcal{B}$ above)
derive from structural properties of $(\mathbb{Z}/N\mathbb{Z})^\times$, and group-theoretical considerations
pertinent to other algorithms may also imply similar benchmarks. The benchmark arising from the
period-finding algorithm is fortuitous.

Further studies may yet show that the benchmarks identified in this paper are toothless. However, the findings
on the period-finding algorithm should still be of interest in view of their generality. 
According to the results in section \ref{sc:asymp}, the approximation $\mathsf{P}^{(\mathfrak{q})}_\infty$ 
has the merit of being a lower bound to the probability of success when $\mathfrak{q}\ge0$ and $1/r$ is negligible.

\bibliographystyle{plain}

\begin{thebibliography}{99}
\bibitem{GKK19} A. Gheorghiu, T. Kapourniotis and E. Kashefi, ``Verification of quantum computation: 
an overview of existing approaches,'' 
\href{https://doi.org/10.1007/s00224-018-9872-3}{Theory Comput. Syst. \textbf{63}, 715--808 (2019)}.

\bibitem{EHW20} J. Eisert, D. Hangleiter, N. Walk, I. Roth,  D. Markham, R. Parekh, U. Chabaud and E. Kashefi, 
``Quantum certification and benchmarking,''
\href{https://doi.org/10.1038/s42254-020-0186-4}{Nat. Rev. Phys. \textbf{2}, 382--90 (2020)}.

\bibitem{KR21} M. Kliesch and I. Roth, ``Theory of quantum system verification,'' 
\href{https://doi.org/10.1103/PRXQuantum.2.010201}{PRX Quantum \textbf{2}, 010201 (2021)}.

\bibitem{Sh97}
P. W. Shor, ``Polynomial time algorithms for prime factorization and discrete logarithms on a
quantum computer,'' \href{https://doi.org/10.1137/S0097539795293172}{SIAM J. Comput.~\textbf{26}(5), 1484--509 (1997)}.

\bibitem{CvD10}
 A. M. Childs and W. van Dam, ``Quantum algorithms for algebraic problems,''
 \href{https://doi.org/10.1103/RevModPhys.82.1}{Rev. Mod. Phys.~\textbf{82}(1), 1--52 (2010)}.
 
\bibitem{DLMF} \href{http://dlmf.nist.gov/}{NIST Digital Library of Mathematical Functions} (Release 1.1.3 of 2021-09-15),
F.~W.~J. Olver, A.~B. {Olde Daalhuis}, D.~W. Lozier, B.~I. Schneider, R.~F. Boisvert, C.~W. Clark, B.~R. Miller, B.~V. Saunders,
H.~S. Cohl, and M.~A. McClain (eds.).

\bibitem{Le02} 
 G. Leander, ``Improving the success probability of Shor's factoring algorithm,'' 
 \href{https://arxiv.org/abs/quant-ph/0208183}{arXiv: quant-ph/0208183}.

 \bibitem{BW07} 
 P. S. Bourdon and H.~T. Williams, ``Sharp probability estimates for Shor's order-finding algorithm,''
 \href{https://doi.org/10.26421/QIC7.5-6-7}{Quantum Information \& Computation~\textbf{7}(5\&6), 522--50 (2007)}.

\bibitem{HW08} 
 G. H. Hardy and E. M. Wright,  ``An introduction to the theory of numbers,'' 
 Oxford University Press, Oxford, UK, 6th ed., 2008.
 
 \bibitem{GE21}
C. Gidney and M. Eker\r{a}, ``How to factor 2048 bit RSA integers in 8 hours using 20 million noisy qubits,''
\href{https://doi.org/10.22331/q-2021-04-15-433}{Quantum \textbf{5}, 433 (2021)}.

\bibitem{GS21}
\'{E} Gouzien and N. Sangouard, ``Factoring 2048-bit RSA integers in 177 days with 13436 qubits and a multimode memory,''
\href{https://doi.org/10.1103/PhysRevLett.127.140503}{Phys. Rev. Lett. \textbf{127}, 140503 (2021)}.
 

\end{thebibliography}

\appendix

\section{Reduction $\Delta_j$ in \eqref{eq:Delta}}\label{app:A}

With the substitution of $2^{m-n_r} j$ in \eqref{eq:Delta} by
\begin{equation}
 \left\lfloor \frac{2^{m-n_r} j}{r_\mathrm{o}}\right\rfloor r_\mathrm{o} + 2^{m-n_r} j (\Mod{r_\mathrm{o}}) ,
\end{equation} 
where it is understood that $2^{m-n_r} j (\Mod{r_\mathrm{o}})$ is the \textit{least\/} non-negative
residue of $2^{m-n_r} j$ modulo $r_\mathrm{o}$ (and, hence, an element of $\mathbb{Z}/r_\mathrm{o}\mathbb{Z}$),
$\Delta_j$ can be rewritten as
\begin{align}\label{eq:deltajalt}
 \Delta_j  & =\left\lfloor \frac{2^{m-n_r} j (\Mod{r_\mathrm{o}})}{r_\mathrm{o}} + \frac{1}{2}\right\rfloor   \\[1ex]
               &  \hspace*{0.15\textwidth}
                       - \frac{2^{m-n_r} j (\Mod{r_\mathrm{o}})}{r_\mathrm{o}} . \nonumber 
\end{align}
Since $r_\mathrm{o}$ is odd, it is coprime to $2^{m-n_r}$, and, just as the integers 
$j\in\mathbb{Z}/r_\mathrm{o}\mathbb{Z}$ form a \textit{complete set of residues\/} modulo $r_\mathrm{o}$, 
so do the $r_\mathrm{o}$ integers $2^{m-n_r}j$  (see, for example, Theorem 56
in \cite{HW08}). The corresponding
least non-negative residues $2^{m-n_r} j (\Mod{r_\mathrm{o}})$ must then be identical to 
$\mathbb{Z}/r_\mathrm{o}\mathbb{Z}$. By an appropriate change of the dummy variable of summation in 
(\ref{eq:reducedPtotal}), it can be arranged that
\begin{equation}\label{eq:DeltaSimple}
 \Delta_j = \left\lfloor  \frac{j}{r_\mathrm{o}} +\frac{1}{2}\right\rfloor - \frac{j}{r_\mathrm{o}}
                       \hspace*{0.05\textwidth} (j\in\mathbb{Z}/r_\mathrm{o}\mathbb{Z}) .
\end{equation}
Evaluation of \eqref{eq:DeltaSimple} yields the following $r_\mathrm{o} - 1$ non-zero values for $\Delta_j$: 
\begin{equation}
    \Delta_j = - \frac{j}{r_\mathrm{o}}
\end{equation}
and
\begin{equation}
  \Delta_{r_\mathrm{o}-j} = + \frac{j}{r_\mathrm{o}}
\end{equation}
for $j\in \{1,2,\ldots, \tfrac{1}{2}(r_\mathrm{o}-1)\}$. As \eqref{eq:DeltaSimple} also trivially
implies that $\Delta_0=0$, the values of $\Delta_j$ in \eqref{eq:DeltaSimple}
clearly coincide with those given in connection with \eqref{eq:BZi}.

\section{Large $r$ expansion of $P_\mathrm{tot}^{(\mathfrak{q})}$} \label{app:B}

For large $r$, the obvious expansion parameter is $1/r$, but there are others which are more convenient
for the treatment of $P_\mathrm{tot}^{(\mathfrak{q})}$. 
Paralleling the derivation 
of \eqref{eq:deltajalt} for $\Delta_j$ in appendix \ref{app:A}, the difference 
\begin{equation}\label{eq:qc3a}
  m_k^{(\mathfrak{q})} - \frac{2^{\mathfrak{m}_\mathfrak{q}}}{r}
  = \left\lfloor\frac{1}{r}(r - 1 - k+ \mathfrak{r} )\right\rfloor   - \frac{\mathfrak{r}}{r} ,
\end{equation}
where $\mathfrak{r}$ is the least non-negative residue of $2^{\mathfrak{m}_\mathfrak{q}}$ modulo $r$.
Inspection of the values that can be attained by the right-hand side of \eqref{eq:qc3a} leads to the conclusion 
that $-(1 - 1/r)\le m_k^{(\mathfrak{q})} - 2^{\mathfrak{m}_\mathfrak{q}}/r\le 1 - 1/r$. (The upper bound is attained when 
$k=0$, $\mathfrak{r}=1$ and the lower bound when $k=r-1=\mathfrak{r}$.)
If $m_k^{(\mathfrak{q})}$ is parametrised as
\begin{equation}\label{eq:qc22}
 m_k^{(\mathfrak{q})} = \frac{2^{\mathfrak{m}_\mathfrak{q}}}{r} ( 1 + \mu) ,
\end{equation}
then the small parameter $\mu$ is such that
\begin{equation}\label{eq:smallmu}
 |\mu|< r/2^{\mathfrak{m}_\mathfrak{q}} < 1/(2^\mathfrak{q}r), 
\end{equation}
where the last inequality relies on
relation \eqref{eq:mqdef} defining $2^{\mathfrak{m}_\mathfrak{q}}$. 
In turn,
\begin{equation}
1/m_k^{(\mathfrak{q})}<(2^\mathfrak{q}r)^{-1} (1+\mu)^{-1}<1/(2^\mathfrak{q}r-1)
\end{equation}
from the reciprocal of \eqref{eq:qc22} and the inequalities
$r/2^{\mathfrak{m}_\mathfrak{q}} < 1/(2^\mathfrak{q}r)$ and $\mu> -(2^\mathfrak{q}r)^{-1}$
which can be read off from \eqref{eq:smallmu}.

On the basis of the expansions
\begin{align}
 \sinc^2& \!\left( \frac{r x}{2^{\mathfrak{m}_\mathfrak{q}}}\right)   \\
 & = 1 - \frac{\pi^2}{3}(1+\mu)^2 \frac{x^2}{ (m_k^{(\mathfrak{q})})^2 } + \ldots , \nonumber \\
\sinc^2&\! \left(\frac{m^{(\mathfrak{q})}_k  r }{2^{\mathfrak{m}_\mathfrak{q}}}x \right) \\
  & = \sinc^2 ( x ) + 2\left[ \sinc(2x)-\sinc^2(x)\right] \mu + \ldots \nonumber
\end{align}
in $1/m_k^{(\mathfrak{q})}$ and $\mu$, respectively, and the identity
$r S_k(0)/2^{ \mathfrak{m}_\mathfrak{q} }=1+\mu$, the leading-order contribution to 
$P_\mathrm{tot}^{(\mathfrak{q})}$
is given by \eqref{eq:domC}, corrections being of order $1/r$.

\section{Expansion of $\mathcal{P}^{(\mathfrak{q})}$ for $\mathfrak{q}=0$} \label{app:C}

The Euler-Maclaurin summation formula (\cite{DLMF}, Eq.~2.10.1) implies that
\begin{align}\label{eq:EMexp}
\mathcal{P}^{(\mathfrak{q})} &= 2 \int\limits_0^{a_\mathfrak{q}}   f( x)\, dx +  f (a_\mathfrak{q}) \frac{1}{r_\mathrm{o}}  
                                                  + \tfrac{1}{6}f^\prime(a_\mathfrak{q})\frac{1}{r_\mathrm{o}^2} \nonumber \\
      &\hspace*{0.05\textwidth} - \tfrac{1}{360} f^{\prime\prime\prime}(a_\mathfrak{q})  \frac{1}{r_\mathrm{o}^4}+ \ldots ,
\end{align}
where $a_\mathfrak{q}=\lfloor 2^\mathfrak{q}r_\mathrm{o}\rfloor/r_\mathrm{o}$, $f(x)=\sinc^2(x)$, and use 
has been of its evenness and the oddness of its odd derivatives.
For $\mathfrak{q}\ge 1$, \eqref{eq:EMexp} is an expansion in inverse powers of $r_\mathrm{o}$ 
as it stands
because $a_\mathfrak{q}\, (=2^{\mathfrak{q}-1})$ is independent of $r_\mathrm{o}$. However,
\begin{equation}
 a_0=\tfrac{1}{2}(1-1/r_\mathrm{o}) ,
\end{equation}
and
\begin{align}
2 \int\limits_0^{a_0}& f(x)\, dx \label{eq:a0i1} \\
 & = \frac{2}{\pi}\mathrm{Si}(\pi - \pi/r_\mathrm{o}) 
                                                                   - (1 - 1/r_\mathrm{o} )f( a_0 ) , \nonumber \\[1ex]
 f(a_0) &= \left(\frac{2}{\pi}\right)^2\left(\frac{\cos\pi/2r_\mathrm{o}}{1-1/r_\mathrm{o}}\right)^2 , \\[1ex]
 f^\prime (a_0) &= \frac{4}{1-1/r_\mathrm{o}}\left[\frac{\sin\pi/r_\mathrm{o}}{\pi(1-1/r_\mathrm{o})}-f(a_0)\right] .
\end{align}
After substitution of \eqref{eq:a0i1} into \eqref{eq:EMexp}, $f(a_\mathrm{o})$ appears in the combination 
$-(1-2/r_\mathrm{o})f(a_\mathrm{o})$ that has the expansion
\begin{equation}\label{eq:fexp}
 -\left(\frac{2}{\pi}\right)^2+\left[1+\left(\frac{2}{\pi}\right)^2\right]\frac{1}{r^2_\mathrm{o}}+ \ldots
\end{equation}
in which terms linear in $1/r_\mathrm{o}$ are absent.
Equation \eqref{eq:asymp0} is obtained when \eqref{eq:fexp} is coupled with use of the expansion
\begin{equation}
\frac{2}{\pi}\mathrm{Si}(\pi-\pi/r_\mathrm{o}) =\frac{2}{\pi}\mathrm{Si}(\pi)-\frac{1}{r_\mathrm{o}^2}+\ldots 
\end{equation}
and replacement of $f^\prime (a_0)$ by $- 4(2/\pi)^2$
in \eqref{eq:EMexp}.

\end{document}